# Subwavelength field localization based on dielectric mesoscale particle with single and blind nanohole array

Oleg V. Minin[a], Song Zhou[b], Yinghui Cao[c], Pavel Baranov[a] and Igor V. Minin[a]
[a]Tomsk Polytechnic University, 30 Lenin Ave., Tomsk 634050 Russia
[b]Jiangsu Key Laboratory of Advanced Manufacturing Technology, Faculty of Mechanical and Material Engineering, Huaiyin Institute of Technology, Huai'an 223003, China.
[c]College of Computer Science and Technology, 2699 Qianjin Street, Jilin University, Changchun 130021, China

## ABSTRACT

Some new unusual physical phenomena and effects associated with dielectric mesoscale particles with Mie size parameter near 10 were studied and have been discovered during the last decade. In this paper, we propose nanoholes structured wavelength-scaled dielectric cubic particle with refractive index near two, where the array of nanoholes can act as a plurality of near-field probes to simultaneously illuminate the sample surface and it has the potential of surpassing the performance of most existing nearfield imaging approaches. We also offer the concept of the single nano-structuring of a dielectric cylinder or sphere made from conventional optical materials. The choice of the diameter of the nanohole in the particle "transfers" it into the resonance mode, when the characteristics of the field localized in the shadow part of the particle are determined not by the wavelength, but by the size of the nanohole. Thus, the diameter of the focused spot at the exit from the particle can be much smaller than the solid immersion diffraction limit.

## 1. INTRODUCTION

The optical microscope lens has a long history from BC. Around the 11th century, plano-convex lenses made of polished beryl were used as reading stones to magnify manuscripts in the Arab world. In the last decade of the 16th century Hans and Zacharias Janssen had invented the microscope by combining two convex lenses. Robert Hooke was the first to publish a fundamental work on microscopy in 1664[1]. It is interesting to note that spherical lenses were used in the very first versions of optical microscopes. In his book Hooke[1] described the manufacture of small spheres by heating the end of thin fiberglass, and later Huygens and van Leeuwenhoek[2] used small spheres which were made by heating a little piece of glass at the point of a needle. The invention of the optical microscope has made it possible for researcher to see small object as the bacteria and blood cells[2], thus has revolutionized the life sciences.

As the old saying goes," Seeing is believing." Therefore, both microscopes and lenses have been improved continuously to increase the optical image quality and resolution. However, the resolution of classical optical microscopes is limited by the loss of evanescent waves, which carry subdiffraction spatial frequency information of the object and decay exponentially with distance from the object. One method of improving the resolution of a microscope is to use dielectric microparticle as superlens[3,4], which converts the evanescent waves that scattered from the object into propagating waves in the far field, forming virtual image that can be captured by a microscope.

The resolution of microparticle superlens is closely related with its focusing properties[3,4]. On one hand, if light beam is incident on microparticles with proper sizes and refractive index, subwavelength focal spots will be generated near the shadow surface of particles, which is referred to as the effect of Photonic NanoJet (PNJ)[4]. On the other hand, if a subwavelength object is located close to the surface of the microparticle superlens, the evanescent waves that scattered by the object will be collected by the microparticle and coupled into the propagating waves, just like the reversal light propagation of PNJ[3-5]. Due to the scattering wave contains both transverse and longitudinal components in the near-fields of PNJ, the focusing spot of microparticle has the elliptical profile, which can also be explained by the Heisenberg's uncertainty principle[6]. Thus, the lateral Full Width at Half Maximum (FWHM) size of the elliptical focal spot of the microparticle can be considered as an indicator for the resolution of the particle-lens.

## 2. SUPERLENS BASED ON DIELECTRIC MESOSCALE CUBIC PARTICLE WITH BLIND NANOHOLE ARRAY

Several methods of further increasing the resolution of microparticles have been considered in the literature[4,5,7]. Among the many methods, we found that under the conditions of the superresonance effect[8-10], which is based on internal partial wave modes of spherical particles and surface wave modes, the field is localized with a lateral resolution of about λ/5.

Recently, it was demonstrated that the size of the focal spot of dielectric mesoscale sphere with refractive index near $n=2$ can be significantly reduced by introducing a nanohole in its shadow surface, which improves the spatial resolution up to λ/40, far beyond the solid immersion diffraction limit $(\lambda/2n)$[11,12]. The field confinement and enhancement effect of the nanohole structured particle is due to the permittivity contrast between the material of the microparticle and the nanohole. It must be mentioned that the new nanophotonic effect under consideration is to some extent similar to the effect of the near field aperture scanning optical microscopy[5], in which evanescent wave illumination is transmitted from a subwavelength aperture at the end of a metal-coated optical fiber. However, the single nanohole structured particle-lens suffers from several limitations, including long scanning time and insufficient contact between shadow surface of particle and object surface.

A schematic diagram for the proposed superlens is plotted in Figure 1(a). The basic structure of the proposed superlens is a closely stacked 5×5 nanohole array at the shadow surface of a dielectric cubic, with edges length $L = \lambda$ and refractive index $n = 2$. The diameter of the nanohole is 0.025λ (λ/40), and the center-to-center distance between adjacent holes is 0.05λ (λ/20). In the simulation, the incident electric field was assumed to be a plane wave which is linearly polarized along the y-axis and propagates along the positive z-axis. The proposed superlens is modeled and simulated by using COMSOL Multiphysic, and the light intensity ($I = E^2$) in the z-y plane and the shadow surface are plotted in Figure 1(b) and 1(c), respectively. Numerical simulation shows, the optical field can be localized inside the cylindrical nanoholes, indicating the proposed structure has the capbility of manipulating and confining light at the nanoscale. The array of nano-holes forms a "patterned illumination" on the surface of the underlying substrate, as shown in Figure 1(c). These illumination light spots are evanescent in nature, containing high-spatial frequency components, can be used as near-field probes to simultaneously examine the sample surface. Figure 1(d) shows the light intensity along a cut line that is parallel to y axis (the polarization direction) and lying above the shadow surface. According to Figure 1(d), the imaging contrast of the light spots at a given spatial frequency is about 0.53, which is defined as $(I_{max}-I_{min})/(I_{max}+I_{min})$, where $I_{max}$ and $I_{min}$ are the maximum and minimum light intensity, respectively. The FWHM sizes of the light spots sizes are ~0.035λ, which is comparable to the nanohole size, and deeply beyond the diffraction limitation of visible light[13-15] and solid immersion criteria[16]. Thus, the proposed nanohole structured microcubic superlens has the potential of surpassing the performance of most existing near-field imaging approaches[4,7] and the array of nanoholes can act as a plurality of near-field probes to simultaneously illuminate the sample surface. Moreover, theoretically, the size of the evanescent wave illumination spots can be controlled by using smaller and denser nanohole array. Our additional simulation showed that, the field localization effect is observed for nanohole array with hole diameter of up to λ/100.

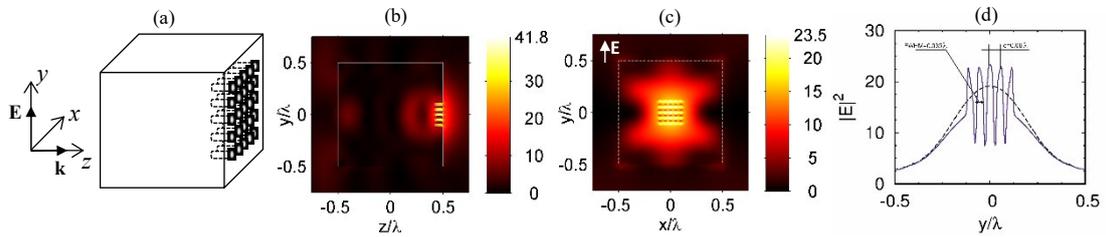

Figure 1. (a) Schematic diagram for the proposed dielectric cubic particle superlens with a 5×5 nanohole array on its shadow surface. Light intensity ($I = E^2$) distributions in (b) z-y plane (side view), (c) the shadow surface (front view), and (d) along a cut line that is parallel to y axis (the polarization direction), and lying just above the shadow surface.

Besides super-resolution imaging, the proposed nanostructued dielectric cubic particle has the potential application of low-cost nanofabrication by producing nano-patterns of high intensity light spots on the substrate. It may also be useful for other applications such as non-linear optical/photonic devices, sensing, energy harvesting, enhanced Raman scattering, optical tweezers[17] and switches[18], hot spots array, etc[19].

## 3. SINGLE PARTICLE WITH HOLE

Recently, it was shown that the size of the field localization area of dielectric wavelength-scaled sphere with refractive index near $n=2$ can be significantly reduced up to $\lambda/100$ by introducing a nanohole in its shadow surface. However common optical materials have a refractive index of about 1.5, so inspired by previous results,[11,12] we decided to extend this effect to the index region of about 1.5. To demonstrate the concept and visualization, we consider a cylindrical particle. The use of cylindrical lenses has a long history[20]. In[21-24] it were shown, that dielectric mesoscale cylindrical particles could be used for improving imaging spatial resolution due to photonic jet effect.

It was found that the use of hole in mesoscale particle increases the resolution dramatically deep below solid immersion criteria[16]. The parameters and geometry of the problem were as follows: cylinder diameter D=3μm, wavelength λ=500 nm, refractive index of the particle material n=1.5. This corresponds to typical parameters in the experimental observation of a photonic jet[25]. To simulate fields localization effects near the boundary of infinite dielectric cylinders, illuminated by a monochromatic plane wave, we use Comsol software (Fig. 2).

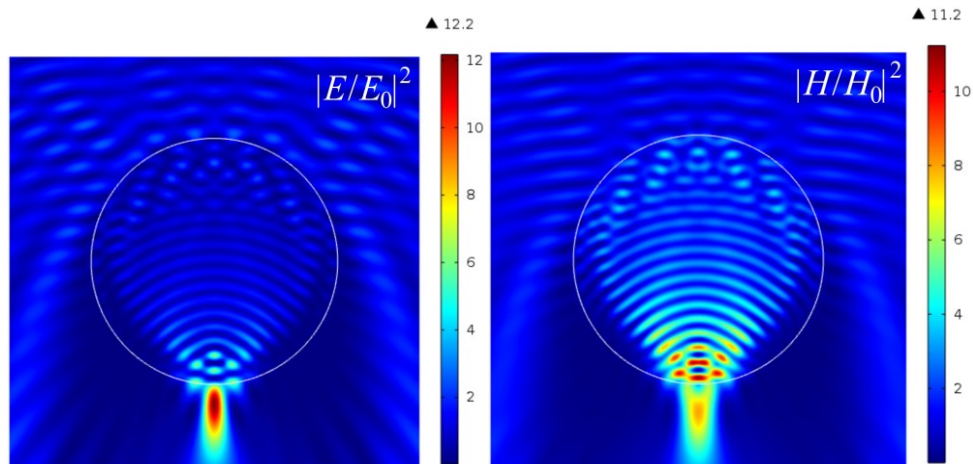

Figure 2. Fields intensity for electric (left) and magnetic (right) components for the dielectric cylindrical particle at nonresonant 500 nm wavelength.

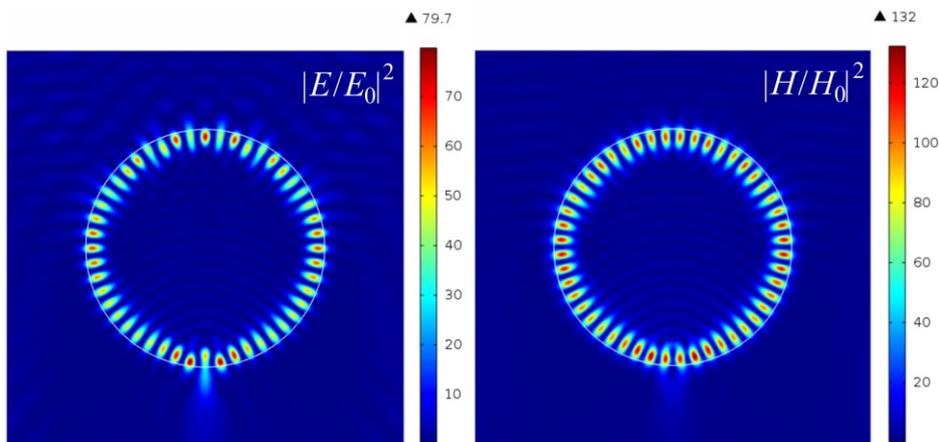

Figure 3. Fields intensity for electric (left) and magnetic (right) components for the dielectric cylindrical particle at resonant 502.3 nm wavelength.

An exact solution of a plane-wave diffraction in an axisymmetric dielectric cylinder was reported in[26]. The resonance radius at which a WGM excited in a dielectric mesoscale cylinder under a plane wave illumination been determinate as the first maximum of the modulus of the Mie series coefficient of cylindrical functions[27,28] and confirmed by Comsol simulations and shown in Fig.3. It could be noted, that the cylindrical wavelength-scaled resonators have a lower Q factor than that for spherical. So, publications dealing with WGM in cylindrical dielectric mesoscale particles are scarce.

The introduction of a defect into a dielectric structure makes it possible to control the degree of fields localization[29]. Changing the nanohole diameter $d_h$ one can find a resonance condition at 500 nm nonresonant wavelength. The spectral shift is due to decrease in the effective refractive index $n_{eff}$ of a nanostructured cylinder: $n_{eff} \approx n(1-V_h/V_0)$, where $V_0$, $V_h$ are the volumes of cylinder and hole, respectively. Taking into account the Mie size parameter $q = \pi D/\lambda_0$ is inversely proportional to the particle refractive index, the resonance wavelength $\lambda_0$ is proportional to $n$[30]. Thus, the decrease in the effective refractive index $n_{eff}$ of the nanohole structured cylinder leads to change in the WGM resonance wavelength $\delta\lambda_0$ according[30] to the relation: $\delta\lambda_0 \propto -(d_h/D)^2$. This effect is shown in Fig.4.

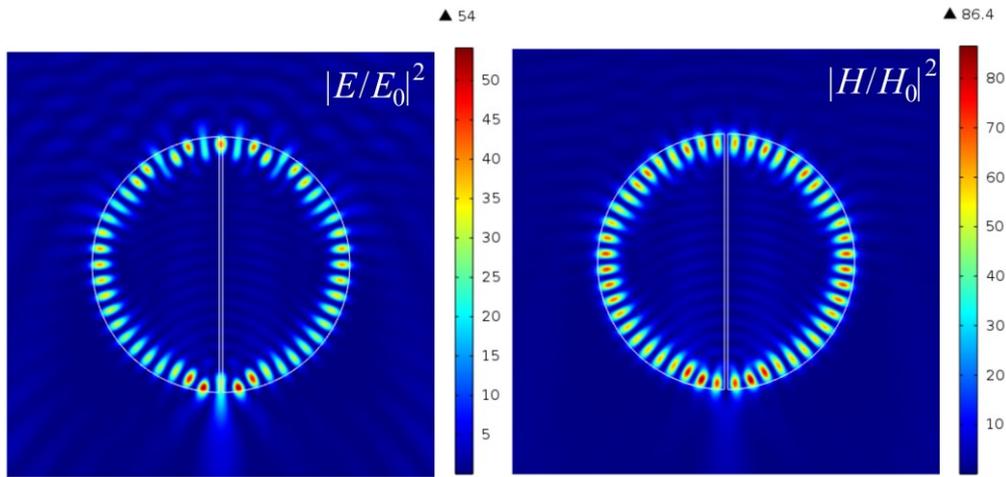

Figure 4. Fields intensity for electric (left) and magnetic (right) components for the dielectric cylindrical particle at non-resonant 500 nm wavelength as in Figure 2 but with through nanohole with diameter of 40 nm.

Figure 5 shows the dependences of the field intensity distribution for the $E^2$ and $H^2$ fields along the photonic jet propagation axis.

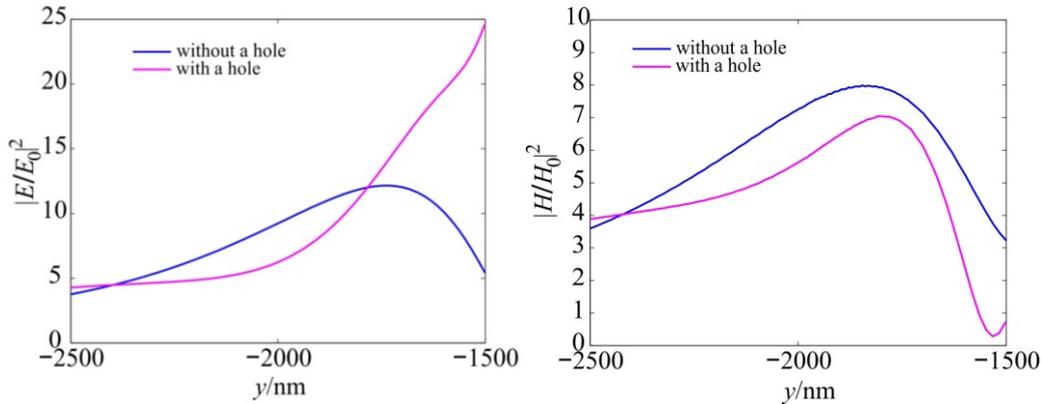

Figure 5. Fields intensity for electric (left) and magnetic (right) components for the dielectric cylindrical particle at non-resonant 500 nm wavelength without/with through nanohole along direction of plane wave under the shadow surface of the particle.

For a nanostructured cylinder, the intensity $E^2$ of the field at the shadow boundary of a particle is higher than for a monolithic one. However, the decrease in intensity with distance is stronger. FWHM of the photonic jet for dielectric cylinder at 500 nm wavelength are: 221 nm (0.442λ) without hole and 113 nm (0.226λ) with 40 nm (0.08λ) hole for $E^2$ fields and 353 nm and 348 nm for $H^2$ field, respectively.

The diameter of the photonic nanojet (FWHM) and the focal distance determines the potential for super-resolution imaging. These parameters vary with the hole diameter in the dielectric particle and refractive index of the particle material. We have shown that the diameter of the photonic nanojets are deep beyond the solid immersion diffraction limit. The results presented only demonstrate the concept of increasing the resolution of a photonic jet regardless of the radiation wavelength and are not the ultimate possible. Optimization of field localization parameters using nanostructuring of cylindrical and spherical particles can find a wide variety of applications[31-41].

## 4. CONCLUSION

Despite the impressive results in the mesoscale dielectric particles phenomenon's[10], the photonics of mesoscale dielectric particles would have a limited scientific interest, if it had no influence on other areas of science. The involvement of additional degrees of freedom in the design of particle- lens, in this case, the nanostructuring of particles in the form of nanoholes, makes it possible to control localized fields in relation to a number of promising practical applications.

## 5. ACKNOWLEDGEMENTS

This work was carried out within the framework by the Tomsk Polytechnic University Development Program. Parts of the work were supported by the Russian Foundation for Basic Research (projects no. 21-57-10001), RSF (project 21-79-00083), Natural Science Research Program of Huai'an (No. HAB202153).